\documentclass[
twocolumn,
]{ceurart}

\usepackage{listings}
\usepackage{caption}
\usepackage{subcaption}
\begin{document}

\copyrightyear{2022}
\copyrightclause{Copyright for this paper by its authors.
  Use permitted under Creative Commons License Attribution 4.0
  International (CC BY 4.0).}

\conference{CIKM'22: Proceedings of the 31st ACM International Conference on Information and Knowledge Management}

\title{Track2Vec: fairness music recommendation with a GPU-free customizable-driven framework}


\author[1]{Wei-Wei Du}[%
orcid=0000-0002-0627-0314,
email=wwdu.cs10@nycu.edu.tw,
url=https://wwweiwei.github.io/,
]
\cormark[1]
\address[1]{Department of Computer Science, National Yang Ming Chiao Tung University, Hsinchu, Taiwan}

\author[1]{Wei-Yao Wang}[%
orcid=0000-0002-6551-1720,
email=sf1638.cs05@nctu.edu.tw,
]

\author[1]{Wen-Chih Peng}[%
orcid=0000-0002-0172-7311,
email=wcpeng@cs.nycu.edu.tw,,
]

\cortext[1]{Corresponding author.}

\begin{abstract}
Recommendation systems have illustrated the significant progress made in characterizing users' preferences based on their past behaviors.
Despite the effectiveness of recommending accurately, there exist several factors that are essential but unexplored for evaluating various facets of recommendation systems, e.g., fairness, diversity, and limited resources.
To address these issues, we propose Track2Vec, a GPU-free customizable-driven framework for fairness music recommendation. 
In order to take both accuracy and fairness into account, our solution consists of three modules, a customized fairness-aware groups for modeling different features based on configurable settings, a track representation learning module for learning better user embedding, and an ensemble module for ranking the recommendation results from different track representation learning modules.
Moreover, inspired by TF-IDF which has been widely used in natural language processing, we introduce a metric called Miss Rate - Inverse Ground Truth Frequency (MR-ITF) to measure the fairness.
Extensive experiments demonstrate that our model achieves a 4th price ranking in a GPU-free environment on the leaderboard in the EvalRS @ CIKM 2022 challenge, which is superior to the official baseline by about 200\% in terms of the official scores.
In addition, the ablation study illustrates the necessity of ensembling each group to acquire both accurate and fair recommendations.
\end{abstract}


\maketitle

\section{Introduction}
Nowadays, there has been a surge in research focusing on recommendation systems (RSs) in different domains (e.g., movies, videos, news, products) with the aim of increasing the possibility of targeting users to view or buy recommended items based on their historical browses.
These approaches introduce their recommendation systems by filtering the most importance and eye-catching information from the collected abundance of data to relieve the information overload problem.
In addition, \citet{covington2016deep} introduced a framework to first select hundreds of video candidates and then rank these videos according to the user history and video content to alleviate the data sparsity problem and to generate more accurate recommendations.


However, most of the work has adopted accuracy-based metrics (e.g., hit-rate (HR), mean reciprocal rank (MRR), and normalized discounted cumulative gain (nDCG)), which fail to consider other factors that reflect the robustness of the models.
Therefore, researchers from both academia and industry have paid more attention to investigating the issues of model fairness and diversity in the past few years.
For instance, \citet{yang2017measuring} introduced fairness measures by generating synthetic data to quantify statistical parity and biases in rankings.
\citet{chia2022beyond} proposed RecList, a general plug-and-play framework to scale up behavioral testing. 

\begin{figure}
    \centering
    \includegraphics[width=0.75\linewidth]{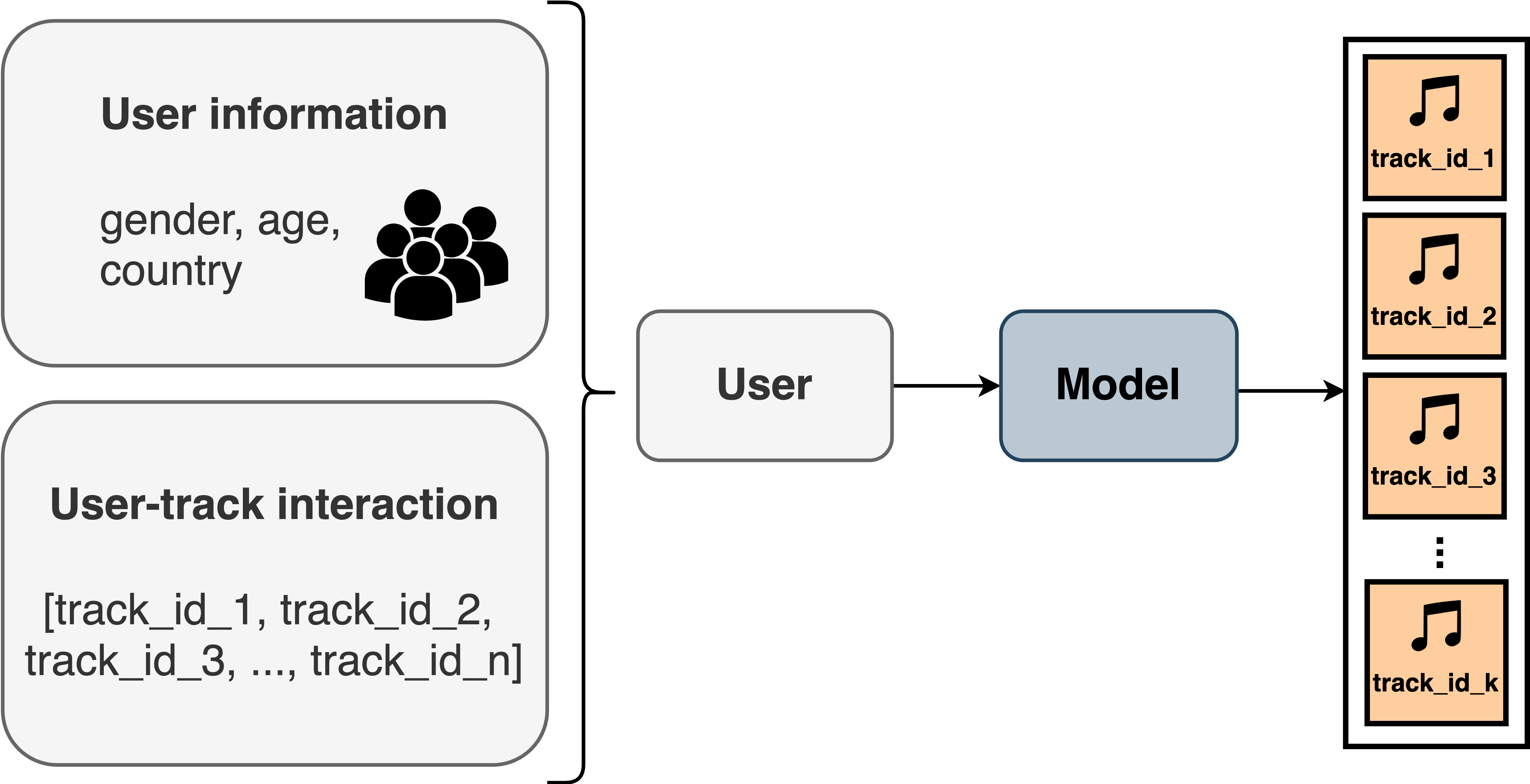}
    \caption{An example of a music recommendation system.}
    \label{fig:example}
\end{figure}

In this challenge hosted by EvalRS\footnote{https://reclist.io/cikm2022-cup/}, given user listening history, track metadata, and user metadata, the goal is to recommend K songs for each user as shown in Figure \ref{fig:example}.
The recommended predictions are evaluated by standard RSs metrics (\textit{HR}, and \textit{MRR}), standard metrics on a per-group or slice basis (\textit{gender balance}, \textit{artist popularity}, \textit{user country}, \textit{song popularity}, and \textit{user history}), and behavioral tests (\textit{be less wrong}, and \textit{latent diversity}) \cite{EvalRS}.
To tackle the shared task, we propose a framework, Track2Vec, a framework with three modules as a fairness music recommendation system.
Specifically, our proposed Track2Vec is composed of a customized fairness-aware groups for dividing user history into multiple facets, a track representation learning module for candidate matching, and an ensemble module for better ranking the recommended results.
In this manner, Track2Vec is able to not only achieve robust performance without auxiliary tasks, but it can also be deployed with limited resources (e.g., a GPU-free machine), which demonstrates the practicality of our framework.

Although there are several metrics that evaluate model behavior with different numbers of bins, we argue that existing metrics that rely on manual settings fail to distinguish the importance of different numbers of classes automatically, which makes it hard to generalize to different tasks and datasets with the same settings.
To that end, we introduce a new metric called Miss Rate - Inverse Ground Truth Frequency (MR-ITF), which computes the less represented categories with more attention to avoiding popular categories dominating the results.
Our proposed metric can be extended to any existing metrics directly by only computing the number of each class as the denominator.
For instance, the numerators can be replaced with MRR or nDCG for evaluating different aspects with the different numbers of classes.

In summary, our contributions are four-fold:
\begin{enumerate}
    \item We propose Track2Vec as a fairness recommendation system with a customizable-driven framework, which achieves effective results (i.e., the fourth prize on the leaderboard) in a GPU-free environment.
    \item To tackle the class imbalance issue, we introduce a customized fairness-aware groups to divide user history into different aspects based on customizable configurations.
    \item We introduce a novel metric, MR-ITF, to measure the predictive distribution of the model by weighting importance based on the number of predictions of each class, which can be generalized to any existing metrics.
    \item We conduct extensive experiments to demonstrate the effectiveness of Track2Vec, which outperforms the official baseline about 200\% in terms of the leaderboard (phase 2) score.
    Moreover, the ablation study verifies the capabilities of the proposed framework. 
\end{enumerate}
\section{Related Work}

\subsection{Recommendation Systems}
Nowadays, recommendation systems are able to solve the information overload problem, which predicts a user's preference based on the user history.
In general, the recommendation techniques can be divided into four categories: content-, collaborative filtering-, knowledge-, and hybrid-based recommendation systems \cite{peng2022survey}.
Recently, deep learning approaches have led to state-of-the-art performance on several recommendation system benchmarks.
For instance, \citet{covington2016deep} introduced a two-stage framework, namely a deep candidate generation model and a deep ranking model, for YouTube recommendation.
PinSage proposed a data efficient Graph Convolutional Network (GCN) algorithm for web-scale recommendation \cite{ying2018graph}.
\citet{grbovic2015commerce}, \citet{vasile2016meta} and \citet{bianchi2020bert} adopted a novel neural language-based algorithm for product recommendation, and \citet{de2021transformers4rec} employed the transformer architecture for session-based recommendation.

In this shared task, one of the constraints is the limited resources and time for training and inferencing.
Therefore, we introduce a data-driven unsupervised approach instead of deep learning supervised approaches to tackle the task with both accurate predictions and efficiency.

\subsection{Fairness Metrics}
Most RSs focus on the accuracy of recommended results, for instance, \cite{ingale2022literature} introduces accuracy, coverage, variety, recommender confidence, robustness, scalability and privacy from a common RSs-centric perspective. 
On the other hand, it is also critical to evaluate trustworthiness, utility, risk and usability from a user-centric perspective of RSs.
In this EvalRS challenge, the organizers provide various metrics for evaluating not only model performance but also model behavior with standard RSs metrics, standard metrics on a per-group or slice basis, and a behavior test.

However, the metrics of model behavior require manual settings of divided bins and are hard to generalize to different tasks.
Thus, we propose a novel metric, MR-ITF, which is a metric that computes frequent categories with lower weights and few categories with higher weights to be sure not to dominate the predictions by majority.
This metric can also be extended to any existing metrics by only modifying the numerators.
\section{Method}

\begin{figure*}
    \centering
    \includegraphics[width=\linewidth]{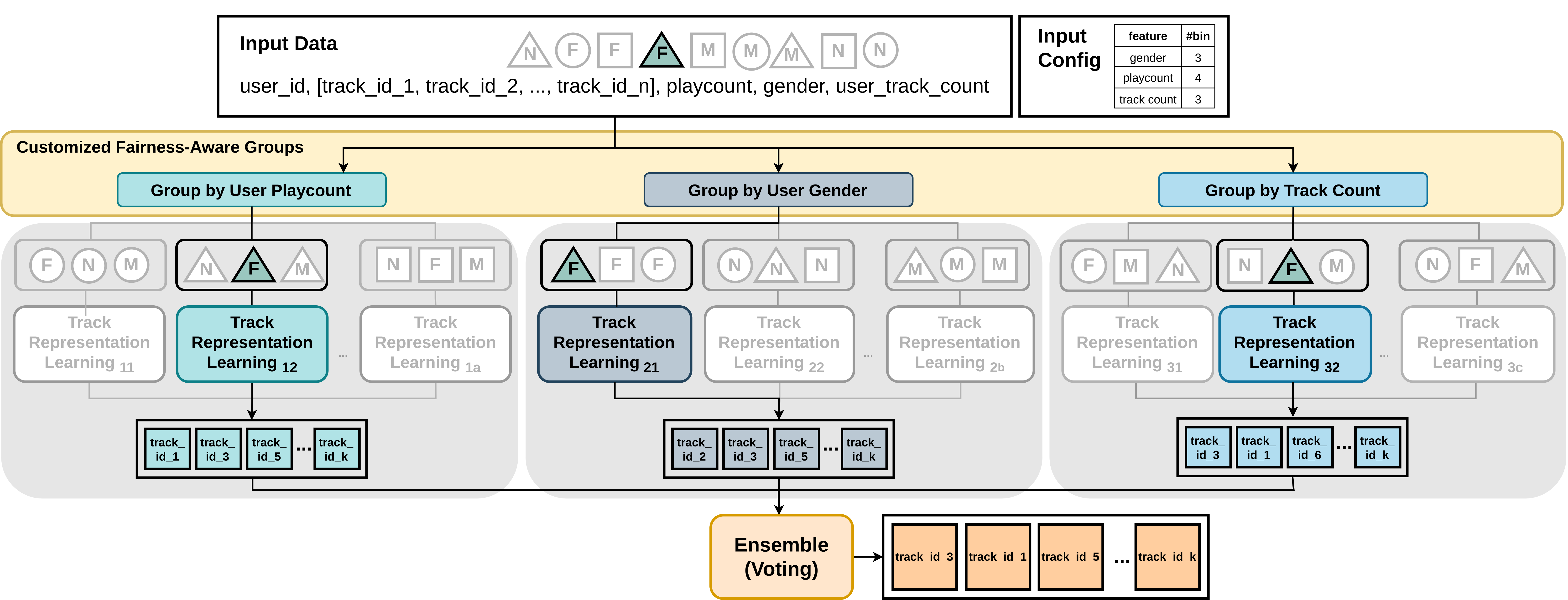}
    \caption{The pipeline of our proposed framework. For every input user sequence (e.g., the green triangle F), our model separates it by three features according to the input configuration (e.g., User Playcount, User Gender and Track Count) and adopts the corresponding track representation learning modules to encode track embeddings for recommend tracks. Then, the outputs from these modules are aggregated by the voting technique to recommend the final recommendations.}
    \label{fig:model}
\end{figure*}

\subsection{Preliminary}
The dataset of this task is based on the LFM-1b Dataset \cite{schedl2016lfm}, corpus of listening events for music recommendation. 
It consists of 100M+ listening events and three types of data, users for user background information and patterns of consumption, tracks which the artist and album belong to, and historical interactions for a collection of interactions between users and tracks.
The details of the data process procedure can be referred to \cite{EvalRS}.

\subsection{Our Recommendation System: Track2Vec}
Figure \ref{fig:model} demonstrates the pipeline of our framework.
Given multiple types of user history, the customized
fairness-aware groups divides each sequence into a different track representation learning module according to the configurable settings. 
For example, if the input configuration is selected to focus on the user, a user sequence will first be checked to be the training instance of the representation learning module in each group (e.g., if the user is male then only the male module has this instance in the gender group).
Afterwards, the predictions are aggregated by using ensemble techniques to generate the top K (K=100 in this paper) predictions for the corresponding user. 
In the training phase, the user features are used to separate user into groups based on the input configuration.
In the testing phase, the user features are obtained by fetching from the training data (by user\_id in this dataset).

\noindent\textbf{Customized Fairness-Aware Groups.}
To enable our model with the fairness behavior, we first discretized each feature based on the feature distribution, and bunched users into different groups by the customizable input configuration to avoid the unbalance issue (e.g., majority dominating the model behavior).

\begin{figure}
     \centering
     \begin{subfigure}[b]{0.15\textwidth}
        \centering
        \includegraphics[width=\linewidth]{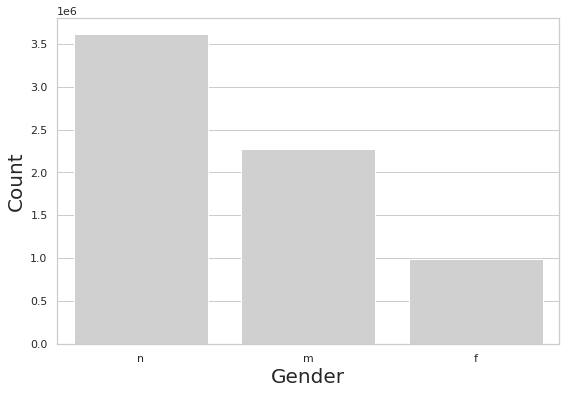}
        \caption{User Gender.}
        \label{fig:gender}
     \end{subfigure}
     \begin{subfigure}[b]{0.15\textwidth}
        \centering
        \includegraphics[width=\linewidth]{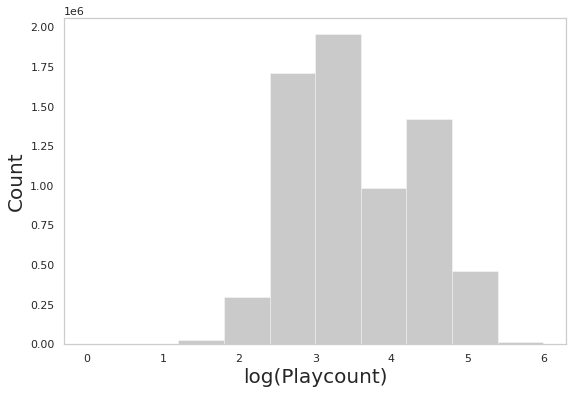}
        \caption{User Playcount.}
        \label{fig:playcount}
     \end{subfigure}
     \begin{subfigure}[b]{0.15\textwidth}
        \centering
        \includegraphics[width=\linewidth]{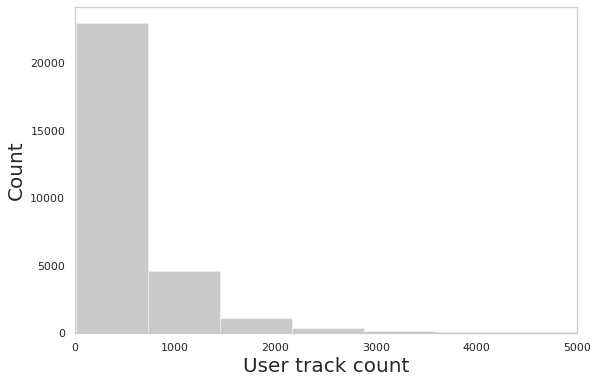}
        \caption{Track Count.}
        \label{fig:utc}
     \end{subfigure}
    \caption{Data distributions of user gender, user playcount, and track count.}
    \label{fig:EDA}
\end{figure}

With the exploratory data analysis shown in Figure \ref{fig:EDA}, we observed that user playcount, user gender and track count are the three most important factors that affect the model behavior metrics. 
The playcount group used logarithmic bucketing in base 10 to divide user into four sub-groups (10, 100, 1000 as divisions in this paper), the gender group divides the each sequence into male, female and neutral, and the track count group also used logarithmic bucketing in base 10 (100, 1000 as division in this paper).
Therefore, we chose these three factors in this work to divide users into the corresponding groups.
We note that these factors are configurable, which can change to others in the framework.

\noindent\textbf{Track Representation Learning.}
One of the limitations in this task is the time constraint (22.5 minutes/fold in average); thus it is challenging to learn a fine-grained track representation using supervised deep learning approaches in a limited amount of time.
Therefore, we focus on an unsupervised method to meet the requirement for training and recommending tracks for users.
Specifically, we employ Word2Vec \cite{mikolov2013efficient} to train track embeddings by calculating the interactions between tracks, which only requires both low computational cost and high-quality.
As there are two options in Word2Vec (i.e., continuous bag-of-words (CBOW) and skip-gram), we experiment using both methods with different negative sampling rates and window sizes to select the best one.
The CBOW architecture predicts the current token based on the whole context, and the skip-gram predicts surrounding tokens given the current token.

\noindent\textbf{Ensemble Techniques.}
Ensembling prediction results have demonstrated the robustness of models in previous work \cite{DBLP:journals/corr/abs-2007-02259,DBLP:conf/aaai/WangP22,DBLP:conf/ltedi/WangTDP22}, which motivated us to adopt ensemble techniques to produce more robust and diverse results.
To consider different factors and to recommend fairer tracks to users, voting is used for ensembling each group with different priorities.
Specifically, the ensemble re-ranking strategy is applied as follows after generating different predictions from each track representation learning module: 
\begin{itemize}
    \item Priority 1: Cumulative recommending times in descending order.
    \item Priority 2: Original individual module ranking in order.
\end{itemize}

\newcommand{\specialcell}[2][c]{%
  \begin{tabular}[#1]{c}#2\end{tabular}}
\begin{table*}
\centering
\caption{\label{table:Table1} Ablation study of Track2Vec. G: Gender. P: Playcount. U: User track count. Total score is computed as ((1) + (2) + (3)) / 3 same as Phase 1 since Phase 2 requires a minimum hit-rate threshold.}
\begin{tabular}{c|c|c|c|c|c|c|c}
    \toprule
     & G & P & U & G+P & G+U & P+U & Track2Vec (ours)\\
    \midrule 
      Standard RSs metrics (1) & 0.0103 & 0.0118 & 0.0128 & 0.0127 & 0.0136 & \underline{0.0143} & \textbf{0.0146} \\
    \midrule
     \specialcell{Standard metrics \\ on a per-group (2)} & -0.0073 & \textbf{-0.0039} & -0.0061 & -0.0052 & -0.0055 &  \underline{-0.0044} & -0.0055 \\
    \midrule
    Behavioral tests (3) & -0.0138 & 0.0014 & 0.0008 & 0.0188 & 0.0156 & \underline{0.0223} & \textbf{0.0271} \\
    \midrule
    MR-ITF (ours) & -4.3862 & -4.3863 & \underline{-4.3861} & -4.3862 & \underline{-4.3861} & \textbf{-4.3860} & \underline{-4.3861}\\
    \bottomrule
    Total Score & -0.0048 & 0.0008 & -0.0003 & 0.0041 & 0.0035 & \underline{0.0057} & \textbf{0.0062} \\
\end{tabular}
\end{table*}



\subsection{Our Fairness Metric: MR-ITF}
Currently, HR, nDCG and MRR are the most used metrics in recommendation systems to evaluate the effectiveness of models, but they fail to reflect the model behavior.
To address the issue, MRED, being less wrong and latent diversity are proposed as an evaluation metric by RecList \cite{chia2022beyond}.
However, these metrics require human settings for the number of bins, but it is hard to generalize the same configurations to other tasks and datasets.
Inspired by term frequency - inverse document frequency \cite{salton1988term}, which is used for considering the frequency of the words and for lowering the importance of the high frequency words, we designed a novel metric, miss rate - inverse ground truth frequency (MR-ITF), to aggregate all scores with different importance weighting for each class.

Formally, the computation of MR-ITF is as follows:
\begin{equation}
    MR-ITF = \frac{\sum_{i=1}^{|C|} MR_i \times ITF_i}{\sum_{j=1}^{N} MR_j} ,
\end{equation}
\begin{equation}
    ITF_i = log (\frac{\# total~ predictions}{\# track_i}),
\end{equation}
where $C$ is the number of classes (the number of tracks in this paper), $N$ is the number of total instances, and $MR_i$ is the miss-rate of the $i$-th track, as in \cite{chia2022beyond}.

If the predictions and ground truths are imbalanced, MR-ITF can attribute more importance to the tracks that are underrepresented. 
That is, MR-ITF relieves the influence of the majority group to dominate the result of whether it is a good model. 
For an edge example with the LFM-1b Dataset, if the number of a less popular song is 1 and the others are all "As It Was", the hit-rate of the model is quite perfect when the predictions are all recommend "As It Was", which is not fair and homogeneous in real-world applications.
In this scenario, MR-ITF can capture this unfair condition to evaluate the model behavior.
It is worth noting that the nominator can be changed to any existing metrics, which not only demonstrates the generalizability of our proposed metric but also the automatic adjustment without manual settings.

\section{Experiment}
\subsection{Experimental Setting}

To implement our Track2Vec, we adopted Word2Vec\cite{rehurek_lrec} as the track representation learning module.
The dimension of each track embedding was set to 100, the window size was set to 60, the minimum track frequency was 0, the number of negative sampling was 5, random seed was set as 27 and the training epochs were set to 10.
All the training and evaluation phases were conducted on a machine with AMD Ryzen Threadripper 3960X 24-Core Processor and 252GB RAM (we do not report our GPU as our approach does not require it).
The results of the ablative experiments is 4-fold boostrapped cross-validation\footnote{Our code will be available at https://github.com/wwweiwei/Track2Vec.}.

\begin{table*}
\small
\centering
\caption{\label{table:Table2} Performance of our Track2Vec in the leaderboard. The formula of the score is normalized with the official baseline and the best score of Phase 1.}
\begin{tabular}{c|c|c|c|c|c}
    \toprule
    Rank & Model & Score & \specialcell{Standard \\RSs metrics} & \specialcell{Standard metrics \\ on a per-group} & Behavioral tests\\
    \midrule
    4 & Track2Vec & 1.1847 & 0.0088 & 2.9481 & 0.2050 \\
    \midrule
    - & CBOWRecSysBaseline & -1.2122 & 0.0512 & -3.7194 & 0.4527 \\
    \bottomrule
        Improvements (\%) & - & \textbf{198} & -83 & \textbf{179} & -55
\end{tabular}
\end{table*}

\subsection{Results}
\noindent\textbf{Offline Performance.}
We first conducted an ablation study to ensure the effective design of our proposed Track2Vec.
As shown in Table \ref{table:Table1}, it is evident that the performance of the total score (i.e., the average of (1) - (3)) is degraded without either groups (gender, playcount, user track count) compared with our Track2Vec.
This result verifies the need to divide a user sequence to the corresponding group based on the configurations.
Moreover, the behavioral tests of G perform the worst, which indicates that using user gender as the splitting standard can achieve better results in each group (i.e., (2)), this hinders the model's recommendation of accurate music tracks to users as well as the fair diversity.
It is noted that the performances of MR-ITF (i.e., our proposed metric) are similar in different models, which indicates that the ground truths of tracks are quite diverse in the test set; thus, the ITR term of each model are nearly the same.
In addition, the miss rates of our variants are similar within different tracks after our investigation.

\noindent\textbf{Testing Performance.}
Table \ref{table:Table2} shows the performance on the EvalRS leaderboard.
Our approach achieved a total score of 1.1847, ranking the fourth prize among 17 teams.
In addition, these results illustrate that our Track2Vec outperforms the official baseline (CBOWRecSysBaseline) by nearly 200\%, which demonstrates the robust capability of our model.
Furthermore, our approach demonstrates that not only using less computation cost in a GPU-free framework, but also utilizing only three features and track\_id can achieve competitive performance.

\begin{table}
\small
\caption{\label{table:Table3} Ensemble groups overlapping ratio.}
\begin{tabular}{c|c}
    \toprule
    Groups & Recommended track\_id \\
    \midrule
    Gender & \specialcell{Rolling in the Deep, Lights, \\ Get Lucky, ...} \\
    \midrule
    Playcount & \specialcell{Burn, Lights, \\ We Found Love, ...} \\
    \midrule
    User Track Count & \specialcell{Set Fire to the Rain, We Found \\ Love,  Titanium, ...} \\
    \midrule
    Track2Vec  & \specialcell{Lights, We Found \\ Love, Burn, ...} \\
    \bottomrule
    Overlapping ratio & 12\% \\
\end{tabular}
\end{table}


\subsection{Case Study: Track2Vec Behavior}
To analyze the behavior of each group, we further conducted a case to demonstrate the overlapping coverage of the top 100 recommended tracks from each group of Track2Vec.
Table \ref{table:Table3} illustrates parts of the recommended results and the overlapping ratio.
We can observe that there is little overlapping of recommendation results across the three groups, which demonstrates the diversity of each group and shows that Track2Vec is capable of considering from these predictions to achieve both accurate and fair recommendations.
\section{Conclusion}
In this paper, we propose Track2Vec as a fairness recommendation system by a customizable—driven groups to achieve fairness model behavior, track representation learning to capture different user preferences and an ensemble technique to aggregate different aspects.
To mitigate the issue of neglecting the minority groups, we introduce MR-ITF by weighting different degrees of importance for each class based on the corresponding frequency, which can be extended to any existing metrics without manual settings.
By conducting extensive experiments, our Track2Vec achieved superior performance compared to the official baseline, which shows not only the capability of recommending fair music tracks but also an efficient recommendation systems without any GPU.
In addition, our proposed MR-ITF is able to reflect prediction bias, which uncovers the model behavior and fosters researchers to develop more advanced systems.

\bibliography{reference}

\begin{thebibliography}{18}
\expandafter\ifx\csname natexlab\endcsname\relax\def\natexlab#1{#1}\fi
\providecommand{\url}[1]{\texttt{#1}}
\providecommand{\href}[2]{#2}
\providecommand{\path}[1]{#1}
\providecommand{\DOIprefix}{doi:}
\providecommand{\ArXivprefix}{arXiv:}
\providecommand{\URLprefix}{URL: }
\providecommand{\Pubmedprefix}{pmid:}
\providecommand{\doi}[1]{\href{http://dx.doi.org/#1}{\path{#1}}}
\providecommand{\Pubmed}[1]{\href{pmid:#1}{\path{#1}}}
\providecommand{\bibinfo}[2]{#2}
\ifx\xfnm\relax \def\xfnm[#1]{\unskip,\space#1}\fi
\bibitem[{Covington et~al.(2016)Covington, Adams, and
  Sargin}]{covington2016deep}
\bibinfo{author}{P.~Covington}, \bibinfo{author}{J.~Adams},
  \bibinfo{author}{E.~Sargin},
\newblock \bibinfo{title}{Deep neural networks for youtube recommendations},
\newblock in: \bibinfo{booktitle}{RecSys}, \bibinfo{publisher}{{ACM}},
  \bibinfo{year}{2016}, pp. \bibinfo{pages}{191--198}.
\bibitem[{Yang and Stoyanovich(2017)}]{yang2017measuring}
\bibinfo{author}{K.~Yang}, \bibinfo{author}{J.~Stoyanovich},
\newblock \bibinfo{title}{Measuring fairness in ranked outputs},
\newblock in: \bibinfo{booktitle}{{SSDBM}}, \bibinfo{publisher}{{ACM}},
  \bibinfo{year}{2017}, pp. \bibinfo{pages}{22:1--22:6}.
\bibitem[{Chia et~al.(2021)Chia, Tagliabue, Bianchi, He, and
  Ko}]{chia2022beyond}
\bibinfo{author}{P.~J. Chia}, \bibinfo{author}{J.~Tagliabue},
  \bibinfo{author}{F.~Bianchi}, \bibinfo{author}{C.~He},
  \bibinfo{author}{B.~Ko},
\newblock \bibinfo{title}{Beyond {NDCG:} behavioral testing of recommender
  systems with reclist},
\newblock \bibinfo{journal}{CoRR} \bibinfo{volume}{abs/2111.09963}
  (\bibinfo{year}{2021}).
\bibitem[{Tagliabue et~al.(2022)Tagliabue, Bianchi, Schnabel, Attanasio, Greco,
  de~Souza P.~Moreira, and Chia}]{EvalRS}
\bibinfo{author}{J.~Tagliabue}, \bibinfo{author}{F.~Bianchi},
  \bibinfo{author}{T.~Schnabel}, \bibinfo{author}{G.~Attanasio},
  \bibinfo{author}{C.~Greco}, \bibinfo{author}{G.~de~Souza P.~Moreira},
  \bibinfo{author}{P.~J. Chia},
\newblock \bibinfo{title}{Evalrs: a rounded evaluation of recommender systems},
\newblock \bibinfo{journal}{CoRR} \bibinfo{volume}{abs/2207.05772}
  (\bibinfo{year}{2022}).
\bibitem[{Peng(2022)}]{peng2022survey}
\bibinfo{author}{Y.~Peng},
\newblock \bibinfo{title}{A survey on modern recommendation system based on big
  data},
\newblock \bibinfo{journal}{CoRR} \bibinfo{volume}{abs/2206.02631}
  (\bibinfo{year}{2022}).
\bibitem[{Ying et~al.(2018)Ying, He, Chen, Eksombatchai, Hamilton, and
  Leskovec}]{ying2018graph}
\bibinfo{author}{R.~Ying}, \bibinfo{author}{R.~He}, \bibinfo{author}{K.~Chen},
  \bibinfo{author}{P.~Eksombatchai}, \bibinfo{author}{W.~L. Hamilton},
  \bibinfo{author}{J.~Leskovec},
\newblock \bibinfo{title}{Graph convolutional neural networks for web-scale
  recommender systems},
\newblock in: \bibinfo{booktitle}{{KDD}}, \bibinfo{publisher}{{ACM}},
  \bibinfo{year}{2018}, pp. \bibinfo{pages}{974--983}.
\bibitem[{Grbovic et~al.(2015)Grbovic, Radosavljevic, Djuric, Bhamidipati,
  Savla, Bhagwan, and Sharp}]{grbovic2015commerce}
\bibinfo{author}{M.~Grbovic}, \bibinfo{author}{V.~Radosavljevic},
  \bibinfo{author}{N.~Djuric}, \bibinfo{author}{N.~Bhamidipati},
  \bibinfo{author}{J.~Savla}, \bibinfo{author}{V.~Bhagwan},
  \bibinfo{author}{D.~Sharp},
\newblock \bibinfo{title}{E-commerce in your inbox: Product recommendations at
  scale},
\newblock in: \bibinfo{booktitle}{Proceedings of the 21th ACM SIGKDD
  international conference on knowledge discovery and data mining},
  \bibinfo{year}{2015}, pp. \bibinfo{pages}{1809--1818}.
\bibitem[{Vasile et~al.(2016)Vasile, Smirnova, and Conneau}]{vasile2016meta}
\bibinfo{author}{F.~Vasile}, \bibinfo{author}{E.~Smirnova},
  \bibinfo{author}{A.~Conneau},
\newblock \bibinfo{title}{Meta-prod2vec: Product embeddings using
  side-information for recommendation},
\newblock in: \bibinfo{booktitle}{Proceedings of the 10th ACM conference on
  recommender systems}, \bibinfo{year}{2016}, pp. \bibinfo{pages}{225--232}.
\bibitem[{Bianchi et~al.(2020)Bianchi, Yu, and Tagliabue}]{bianchi2020bert}
\bibinfo{author}{F.~Bianchi}, \bibinfo{author}{B.~Yu},
  \bibinfo{author}{J.~Tagliabue},
\newblock \bibinfo{title}{Bert goes shopping: Comparing distributional models
  for product representations},
\newblock \bibinfo{journal}{arXiv preprint arXiv:2012.09807}
  (\bibinfo{year}{2020}).
\bibitem[{de~Souza Pereira~Moreira et~al.(2021)de~Souza Pereira~Moreira, Rabhi,
  Lee, Ak, and Oldridge}]{de2021transformers4rec}
\bibinfo{author}{G.~de~Souza Pereira~Moreira}, \bibinfo{author}{S.~Rabhi},
  \bibinfo{author}{J.~M. Lee}, \bibinfo{author}{R.~Ak},
  \bibinfo{author}{E.~Oldridge},
\newblock \bibinfo{title}{Transformers4rec: Bridging the gap between nlp and
  sequential/session-based recommendation},
\newblock in: \bibinfo{booktitle}{Fifteenth ACM Conference on Recommender
  Systems}, \bibinfo{year}{2021}, pp. \bibinfo{pages}{143--153}.
\bibitem[{Ingale and Ellambotla(2022)}]{ingale2022literature}
\bibinfo{author}{V.~Ingale}, \bibinfo{author}{S.~Ellambotla},
\newblock \bibinfo{title}{Literature review on performance evaluation of
  recommendation system with different dimensions of metrics},
\newblock \bibinfo{journal}{Available at SSRN 4140551}  (\bibinfo{year}{2022}).
\bibitem[{Schedl(2016)}]{schedl2016lfm}
\bibinfo{author}{M.~Schedl},
\newblock \bibinfo{title}{The lfm-1b dataset for music retrieval and
  recommendation},
\newblock in: \bibinfo{booktitle}{{ICMR}}, \bibinfo{publisher}{{ACM}},
  \bibinfo{year}{2016}, pp. \bibinfo{pages}{103--110}.
\bibitem[{Mikolov et~al.(2013)Mikolov, Chen, Corrado, and
  Dean}]{mikolov2013efficient}
\bibinfo{author}{T.~Mikolov}, \bibinfo{author}{K.~Chen},
  \bibinfo{author}{G.~Corrado}, \bibinfo{author}{J.~Dean},
\newblock \bibinfo{title}{Efficient estimation of word representations in
  vector space},
\newblock in: \bibinfo{booktitle}{{ICLR} (Workshop Poster)},
  \bibinfo{year}{2013}.
\bibitem[{Wang et~al.(2020)Wang, Chang, and
  Tang}]{DBLP:journals/corr/abs-2007-02259}
\bibinfo{author}{W.~Wang}, \bibinfo{author}{K.~Chang},
  \bibinfo{author}{Y.~Tang},
\newblock \bibinfo{title}{Emotiongif-yankee: {A} sentiment classifier with
  robust model based ensemble methods},
\newblock \bibinfo{journal}{CoRR} \bibinfo{volume}{abs/2007.02259}
  (\bibinfo{year}{2020}).
\bibitem[{Wang and Peng(2022)}]{DBLP:conf/aaai/WangP22}
\bibinfo{author}{W.~Wang}, \bibinfo{author}{W.~Peng},
\newblock \bibinfo{title}{Team yao at factify 2022: Utilizing pre-trained
  models and co-attention networks for multi-modal fact verification (short
  paper)},
\newblock in: \bibinfo{booktitle}{DE-FACTIFY@AAAI}, volume
  \bibinfo{volume}{3199} of \textit{\bibinfo{series}{{CEUR} Workshop
  Proceedings}}, \bibinfo{publisher}{CEUR-WS.org}, \bibinfo{year}{2022}.
\bibitem[{Wang et~al.(2022)Wang, Tang, Du, and
  Peng}]{DBLP:conf/ltedi/WangTDP22}
\bibinfo{author}{W.~Wang}, \bibinfo{author}{Y.~Tang}, \bibinfo{author}{W.~Du},
  \bibinfo{author}{W.~Peng},
\newblock \bibinfo{title}{Nycu{\_}twd@lt-edi-acl2022: Ensemble models with
  {VADER} and contrastive learning for detecting signs of depression from
  social media},
\newblock in: \bibinfo{booktitle}{{LT-EDI}}, \bibinfo{publisher}{Association
  for Computational Linguistics}, \bibinfo{year}{2022}, pp.
  \bibinfo{pages}{136--139}.
\bibitem[{Salton and Buckley(1988)}]{salton1988term}
\bibinfo{author}{G.~Salton}, \bibinfo{author}{C.~Buckley},
\newblock \bibinfo{title}{Term-weighting approaches in automatic text
  retrieval},
\newblock \bibinfo{journal}{Inf. Process. Manag.} \bibinfo{volume}{24}
  (\bibinfo{year}{1988}) \bibinfo{pages}{513--523}.
\bibitem[{{\v R}eh{\r u}{\v r}ek and Sojka(2010)}]{rehurek_lrec}
\bibinfo{author}{R.~{\v R}eh{\r u}{\v r}ek}, \bibinfo{author}{P.~Sojka},
\newblock \bibinfo{title}{{Software Framework for Topic Modelling with Large
  Corpora}},
\newblock in: \bibinfo{booktitle}{{Proceedings of the LREC 2010 Workshop on New
  Challenges for NLP Frameworks}}, \bibinfo{publisher}{ELRA},
  \bibinfo{address}{Valletta, Malta}, \bibinfo{year}{2010}, pp.
  \bibinfo{pages}{45--50}.
  \bibinfo{note}{\url{http://is.muni.cz/publication/884893/en}}.

\end{thebibliography}

\end{document}